%% file: main.tex
\pdfoutput=1
\PassOptionsToPackage{bookmarksnumbered,unicode}{hyperref}
\RequirePackage{hyperref}

\documentclass[sigconf,9pt]{acmart}

\usepackage{caption}
\usepackage{setspace}
\usepackage{enumitem}

\usepackage{subcaption}
\usepackage{bbm}
\usepackage[ruled,linesnumbered]{algorithm2e}
\usepackage{tabularx}
\usepackage{multirow}

\usepackage{booktabs,subcaption,amsfonts,dcolumn}
\newcolumntype{d}[1]{D..{#1}}

\usepackage{diagbox}
\usepackage{flushend}

\usepackage{multirow}
\usepackage{multicol}

\usepackage{mathtools}

\newlength{\myeqskip}  
\AtBeginDocument{%
    \setlength\abovedisplayskip{\myeqskip}%
    \setlength\belowdisplayskip{\myeqskip}%
    \setlength\abovedisplayshortskip{\myeqskip-\baselineskip}%
    \setlength\belowdisplayshortskip{\myeqskip}}

\acmYear{2025}\copyrightyear{2025}
\setcopyright{rightsretained}
\acmConference[MobiSys '25]{The 23rd Annual International Conference on Mobile Systems, Applications and Services}{June 23--27, 2025}{Anaheim, CA, USA}
\acmBooktitle{The 23rd Annual International Conference on Mobile Systems, Applications and Services (MobiSys '25), June 23--27, 2025, Anaheim, CA, USA}
\acmDOI{10.1145/3711875.3729145}
\acmISBN{979-8-4007-1453-5/25/06}

\begin{document}

\title[DAF]{DAF: An Efficient End-to-End Dynamic Activation Framework for on-Device DNN Training }


\author{Renyuan Liu}
\affiliation{
  \institution{George Mason University}
  \city{Fairfax}
  \state{VA}
  \country{USA}
  \postcode{22030}
}
\email{rliu23@gmu.edu}

\author{Yuyang Leng}
\affiliation{
  \institution{George Mason University}
  \city{Fairfax}
  \state{VA}
  \country{USA}
  \postcode{22030}
}
\email{yleng2@gmu.edu}

\author{Kaiyan Liu}
\affiliation{
  \institution{George Mason University}
  \city{Fairfax}
  \state{VA}
  \country{USA}
  \postcode{22030}
}
\email{kliu23@gmu.edu}

\author{Shaohan Hu}
\affiliation{
  \institution{Global Technology Applied Research, JPMorganChase}
  \city{New York}
  \state{NY}
  \country{USA}
  \postcode{10017}
}
\email{shaohan.hu@jpmchase.com}

\author{Chun-Fu (Richard) Chen}
\affiliation{
  \institution{Global Technology Applied Research, JPMorganChase}
  \city{New York}
  \state{NY}
  \country{USA}
  \postcode{10017}
}
\email{richard.cf.chen@jpmchase.com}

\author{Peijun Zhao}
\affiliation{
  \institution{Global Technology Applied Research, JPMorganChase}
  \city{New York}
  \state{NY}
  \country{USA}
  \postcode{10017}
}
\email{peijun.zhao@jpmchase.com}

\author{Heechul Yun}
\affiliation{
  \institution{University of Kansas}
  \city{Lawrence}
  \state{KS}
  \country{USA}
  \postcode{66045}
}
\email{heechul.yun@ku.edu}

\author{Shuochao Yao}
\affiliation{
  \institution{George Mason University}
  \city{Fairfax}
  \state{VA}
  \country{USA}
  \postcode{22030}
}
\email{shuochao@gmu.edu}

\renewcommand{\shortauthors}{R. Liu et al.}

\sloppy

\input{Abstract}



\maketitle

{

\input{1-Introduction}

\input{2-Motivation}

\input{3-Model}

\input{4-Evaluation}

\input{5-Future_Work}
\input{6-Conclusion}
\input{7-DISCLAIMER}

\section{Acknowledgements}
This work is in part supported by the National Science Foundation grants IIS-2107200, CNS-2038658, and CNS-2038923.

}

\newpage
 \balance
\bibliographystyle{unsrtnat}
\bibliography{references}
\newpage

\end{document}

%% file: Abstract.tex
\begin{abstract}

Recent advancements in on-device training for deep neural networks have underscored the critical need for efficient activation compression to overcome the memory constraints of mobile and edge devices. As activations dominate memory usage during training and are essential for gradient computation, compressing them without compromising accuracy remains a key research challenge.
While existing methods for dynamic activation quantization promise theoretical memory savings, their practical deployment is impeded by system-level challenges such as computational overhead and memory fragmentation.

To address these challenges, we introduce DAF, a Dynamic Activation Framework that enables scalable and efficient on-device training through system-level optimizations. DAF achieves both memory- and time-efficient dynamic quantization training by addressing key system bottlenecks. It develops hybrid reduction operations tailored to the memory hierarchies of mobile and edge SoCs, leverages collaborative CPU-GPU bit-packing for efficient dynamic quantization, and implements an importance-aware paging memory management scheme to reduce fragmentation and support dynamic memory adjustments.
These optimizations collectively enable DAF to achieve substantial memory savings and speedup without compromising model training accuracy.
Evaluations on various deep learning models across embedded and mobile platforms demonstrate up to a $22.9\times$ reduction in memory usage and a $3.2\times$ speedup, making DAF a scalable and practical solution for resource-constrained environments.

\end{abstract}

%% file: 1-Introduction.tex
\section{Introduction}\label{sec:introduction}

In recent years, the adoption of deep neural networks (DNNs) has grown rapidly on mobile and edge devices, enabling a wide range of applications such as autonomous driving~\cite{he2023vi,zhang2023robust,zheng2023autofed, scaleflow}, personal assistance~\cite{benazir2024speech,wen2024autodroid,10.1145/3636534.3690682}, augmented reality~\cite{10.1145/3636534.3690676,kong2024arise,jeong2022band,wu2024theia}, and intelligent sensing~\cite{zheng2019zero,chen2023rf,ouyang2022cosmo}.
This rapid growth has sparked increasing interest in enabling DNN training and fine-tuning directly on users' devices, motivated by increasing concerns over data privacy and the demand for personalized models~\cite{privacy,lin2022device,qiu2022zerofl,gao2018spotlight}. 
Despite its potential, on-device training faces significant challenges, with limited memory capacity on mobile and edge devices being one of the most critical.
A substantial portion of memory consumption during training arises from storing activations, the intermediate outputs of neural network layers needed for calculating weight gradients during backpropagation. To address this bottleneck, compressing activations has become a promising strategy to reduce memory usage.

Numerous studies have explored methods to reduce memory usage~\cite{actnn,wang2023division,koster2017flexpoint,fu2020fractrain,gact,pact,fixedpoint,dynaspa,gim2022memory,poet,elastictrainer}. 
Among these, quantization stands out as a widely used technique. However, fixed bit-width quantization often proves suboptimal due to the varying importance of activations: some can be aggressively quantized or even dropped with minimal impact, while others require higher precision to preserve accuracy. As a result, using a uniform bit-width either restricts the potential compression ratio or significantly compromises accuracy.
To address this challenge, researchers have explored dynamic activation compression techniques~\cite{actnn,gact,fu2020fractrain,elastictrainer}, which adaptively adjust the bit-width of activations during training. Skipping the storage of activations for certain layers can be seen as a special case of quantization, where the bit-width is effectively reduced to zero. These dynamic approaches significantly improve the theoretical memory compression ratio, achieving a good tradeoff between memory efficiency and training accuracy.

While existing studies primarily focus on algorithmic improvements, they often emphasize achieving theoretically optimal compression ratios through simulated quantization. However, in practical scenarios, these approaches face two major drawbacks: the actual memory savings often fall short of theoretical predictions, and the training process introduces significant time overhead, making these methods less practical for real-world deployment.
Through our investigation of real-world training scenarios, we identified three critical challenges that demand immediate attention:
\begin{itemize}[nosep,leftmargin=*]
\item \textbf{Time Overhead of Dynamic Quantization:} Dynamic quantization introduces significant time overhead, particularly during collective reduction operations that require traversing every value in a tensor. This memory-bound operation is notably slow on edge and mobile devices. While NVIDIA's CUB library~\cite{CUB} offers optimizations to accelerate such tasks, these are primarily designed for server GPUs and do not account for the slower memory speeds of mobile platforms. As a result, this frequently used operation in dynamic activation quantization becomes a major performance bottleneck.
\item \textbf{Efficient Handling of Dynamic Bit-Width Layouts:} Effectively managing dynamic bit-width layout conversions presents another significant challenge. Designing a system to store and retrieve quantized activations in a highly compact format is crucial but complex. The difficulty lies in achieving this without incurring excessive memory usage or computational overhead, which could offset the benefits of quantization.
\item \textbf{Memory Management and Fragmentation:} Frequent dynamic quantization, deletion, and storage operations can lead to memory fragmentation, a challenge often overlooked in existing studies, which focus primarily on compression efficiency~\cite{elastictrainer,fu2020fractrain}. Some prior works address fragmentation by periodically invoking empty\_cache calls in the training process~\cite{actnn,gact}. When memory fragmentation or activation usage exceeds a threshold, these mechanisms clear PyTorch's memory pool to reclaim fragmented memory. However, such solutions are extremely inefficient and introduce additional time overhead, further complicating their practical application.
\end{itemize}

To address these challenges in on-device training, we propose \textbf{DAF}, a \textit{Dynamic Activation Framework}. DAF offers a practical, end-to-end solution by incorporating system-level optimizations to effectively mitigate these bottlenecks, all while ensuring negligible accuracy degradation.

Specifically, DAF introduces three key optimizations. 
First, to accelerate collective reduction operations on different memory hierarchies, DAF introduces two complementary methods: parallel reduction and atomic reduction. By seamlessly integrating these methods at both block and thread levels, DAF can select the most efficient reduction strategy tailored to the memory hierarchy of the target platform and task, ensuring optimal performance.

Second, bit layout conversion, where quantized values are transformed into lower-bit integer formats and packed together to save memory footprint, can become a performance bottleneck. DAF addresses this issue by enabling collaborative packing and unpacking between the CPU and GPU. To resolve mismatches in packing layouts between the two, DAF introduces a unified layout: it aligns access logic with GPU gradient computation tile blocks while keeping the low-level packing consistent with the CPU layout. This unified layout ensures that packing and unpacking processes are both memory-efficient and energy-efficient.

Third, to address memory management challenges, DAF employs a static memory allocation strategy combined with page-based management to dynamically add, remove, and quantize activations. For real-time dynamic quantization decisions, DAF utilizes a red-black tree to manage memory pages, enabling fast indexing and efficient removal of activations. Additionally, to accommodate the fluctuating memory budget demands of mobile platforms, DAF can adaptively adjust the size of activation storage in response to budget changes, ensuring optimal resource utilization.

\begin{figure}[!t]
\centering
{\includegraphics[width=0.98\linewidth]{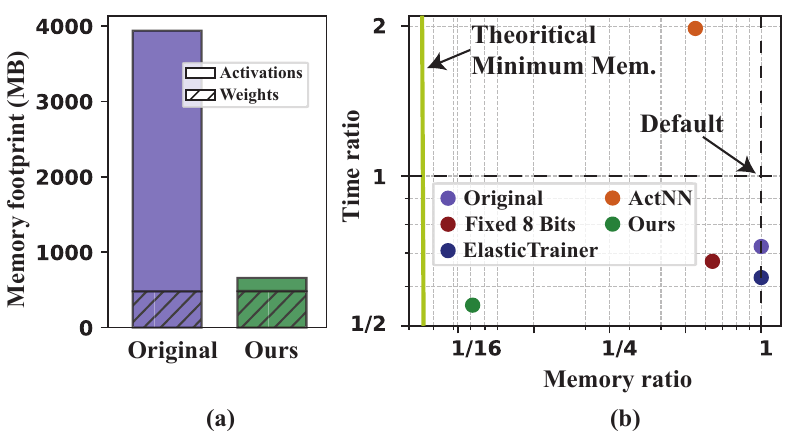}}
\caption{\small (a) Memory footprint during RoBERTa-base LoRA training. (b) Comparison of time and memory usage for various methods under a 1\% accuracy drop, relative to the default setup. \textbf{Original:} Default setup with 1\% accuracy drop; \textbf{Fixed 8 Bits:} All weights and activations quantized to fixed 8 bits; \textbf{ElasticTrainer~\cite{elastictrainer}} and \textbf{ActNN~\cite{actnn}:} Different activation compression training algorithms.}
\label{fig:motivation}
\end{figure}

We evaluate DAF on a convolution-based model, ResNet-18~\cite{resnet}, and two transformer-based models, RoBERTa-base~\cite{roberta} and GPT-2 Medium~\cite{gpt2}, across two embedded platforms (Jetson AGX Orin and Jetson AGX Xavier) and two mobile devices (iQOO Neo 3 and RedMagic 7). Compared to baseline training methods, DAF delivers up to $22.9\times$ lower memory usage and up to $3.2\times$ faster training speed.

In summary, DAF integrates both time and memory optimization strategies to provide a scalable solution for reducing memory usage and time overhead during on-device training, without compromising training accuracy. Our paper makes the following contributions:

\begin{itemize}[nosep,leftmargin=*]
    \item \textbf{DAF Framework:} We propose a comprehensive end-to-end framework that leverages system-level optimizations to drastically reduce memory usage and time overhead during on-device training.
    \item \textbf{Dynamic Quantization Optimization:} We present and integrate two complementary strategies to accelerate frequently used collective reduction operations across a variety of edge and mobile platforms.
    \item \textbf{Bit Layout Conversion Optimization:} We design a collaborative CPU-GPU approach with bit layout conversion, enabling dynamic quantization to achieve both exceptional memory efficiency and time efficiency.
    \item \textbf{Memory Management Optimization:} We design and implement an importance-aware paging memory management scheme to address fragmentation issue caused by dynamic activation quantization. 
    \item \textbf{Practical Memory Reduction and Speedup:} We evaluate DAF across diverse DNN models, tasks, and hardware platforms, demonstrating significant system improvements over STOAs.
\end{itemize}

%% file: 2-Motivation.tex
\section{Background \& Motivation}
\subsection{Dynamic Activation Quantization Training}
Activations are the intermediate outputs generated by each layer of a neural network during the forward pass, temporarily stored to enable gradient computation in the backward pass.
Due to their substantial memory footprint, activations often dominate memory usage during training, making their efficient management a critical challenge, particularly on resource-constrained edge devices. For example, as shown in Figure~\ref{fig:motivation} (a), activations account for 84\% of total memory usage when fine-tuning RoBERTa-base with LoRA~\cite{lora}.

To address the memory constraints of edge and mobile devices, more memory-efficient training strategy must be developed to accommodate various models. A particularly effective approach is the quantization of activations. Since these stored activations are essential for gradient computation, minimizing the impact of quantization on model accuracy is crucial. To this end, many studies have explored dynamic activation quantization, which adapts the bit width of activations for each operation based on specific criteria~\cite{actnn,gact,fu2020fractrain}. 
This method, in theory, can significantly reduce the memory required for on-device training while maintaining a negligible effect on final accuracy.

\begin{figure}[!t]
\centering
{\includegraphics[width=0.98\linewidth]{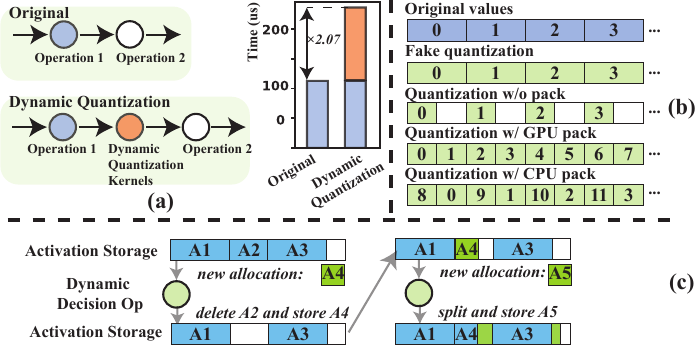}}
\vspace{-0.2cm}
\caption{\small An illustration of practical gaps for dynamic activation quantization training. (a) Runtime overhead due to additional computation. (b) Different bit package strategies. (c) Page management system.}
\label{fig:motivation_3_keys}
\vspace{0.1cm}
\end{figure}

\subsection{Practical Gaps for Dynamic Activation Quantization Training}

\begin{figure*}[!htb]
\centering
    {\includegraphics[width=0.85\linewidth]{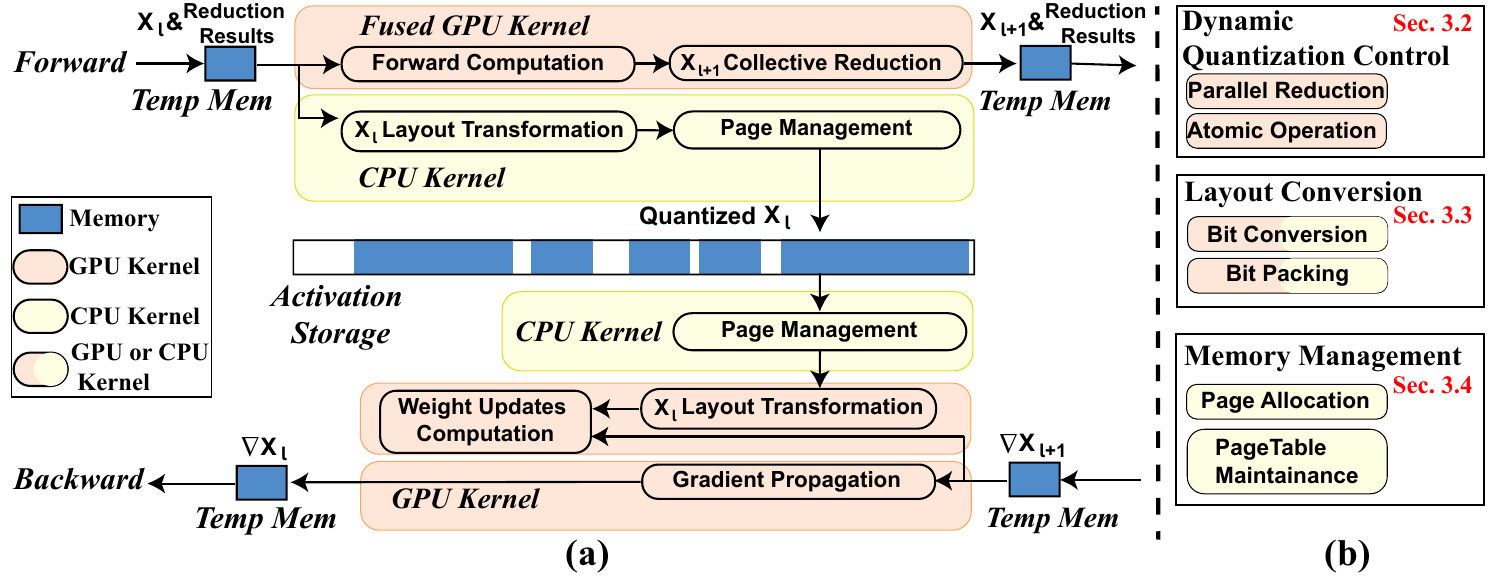}
    \vspace{-0.35cm}
}
         \caption{\small DAF overview. (a) Default DAF computation flow. (b) DAF library.
         }
         \vspace{-0.5cm}
         \label{fig:overview}
\end{figure*}

Existing dynamic activation quantization training methods often fail to achieve their theoretical memory efficiency in practice. The actual memory usage is significantly higher than the theoretical expectation, and the complexity of quantization calculations, and the added complexity of quantization computations, including determining appropriate bit widths, further prolongs training time (as shown in Figure~\ref{fig:motivation}(b)).
These practical inefficiencies stem from three primary challenges: dynamic quantization control, bit layout conversion, and effective memory management.

\textbf{Dynamic Quantization Control.}  
In dynamic activation quantization, each activation is dynamically processed to determine its bit width and compute its quantized value, one of the most frequently executed operations during training.
However, as illustrated in Figure~\ref{fig:motivation_3_keys} (a), it can introduce up to twice the overhead of the original operations. 
The most time-consuming component of this process is the collective reduction operation.
While NVIDIA CUB~\cite{CUB} library provides optimized reduction operations for server GPUs, its performance on mobile and edge GPUs is suboptimal. 
The ratio between memory access speeds across memory hierarchy and computation speeds varies significantly. This discrepancy between mobile and server GPUs makes memory-bound reduction operations less efficient, even with CUB optimizations. 
We propose two complementary strategies, parallel and atomic reduction, to enable efficient collective reduction operations across various memory levels.

\textbf{Bit Layout Conversion.}
Many existing methods focus solely on calculating the theoretically compressible bit width but fail to implement practical storage optimizations. For example, activations are often stored in \texttt{int8} format even when they could be compressed to 2 bits. Achieving efficient quantization requires not only compressing activations to their target bit widths but also converting formats and compactly packing multiple compressed bits together, both of which pose significant technical challenges.
To enable more compact packing and reduce time overhead, we leverage the unified memory architecture of SoCs, enabling collaboration between the CPU and GPU for packing and unpacking tasks. 
However, the parallelism characteristics of the CPU and GPU differ, leading to distinct optimal data processing layouts for these operations, as shown in Figure~\ref{fig:motivation_3_keys} (b).
Identifying an appropriate trade-off to harmonize CPU-GPU cooperation is critical for maximizing efficiency.

\textbf{Memory Management.}  
Frequent dynamic quantization computations, deletions, and insertion can lead to severe memory fragmentation. Some studies~\cite{actnn,gact} attempt to address this issue by manually invoking empty\_cache function, but the overall memory compression ratio remains low, as indicated by the orange points in Figure~\ref{fig:motivation} (b). Similarly, TensorFlow Lite’s static graph approach fails to solve this problem because the size of each activation is dynamic, making it impossible to preallocate storage.
To tackle this challenge, we propose preallocating a contiguous memory block and employing a page-based management scheme for dynamically storing and eliminating activations. This approach combines the benefits of low fragmentation from static graphs with the flexibility of dynamic allocation. However, to support iterative dynamic insertion and deletion of activations, we need a mechanism for fast indexing. To address this, we introduce a CPU-assisted page management system that leverages a red-black tree to efficiently organize and manage activations.




\subsection{Quantization Calculation}\label{sec:preliminary}


Quantization involves mapping a range of values to a limited set of levels, typically defined by a bit-width \( b \). For instance, the quantization process for an activation is governed by the following equations:

\begin{equation}
    \text{scale} = \frac{\text{max} - \text{min}}{2^b - 1}
    \quad 
    \mathbf{Q} = \text{round}\left(\frac{\mathbf{X} - \text{min}}{\text{scale}}\right)
    \label{eq:quant}
\end{equation}




Here, \( b \) is the bit-width used for quantization, while \( \text{min} \) and \( \text{max} \) represent the minimum and maximum values, respectively. The quantized tensor \( \mathbf{Q} \) is obtained by applying these parameters to map the original tensor \( \mathbf{X} \) into a discrete space defined by the bit-width.

The quantization process of DAF does not alter the forward pass computation results. During the forward pass, activations are computed as usual, then quantized and stored in memory. During the subsequent backward pass, these stored activations are dequantized to their original precision for gradient computation. The gradient propagation through layer \( l \) is governed by:

\begin{equation}
\nabla \mathbf{X}_l = \nabla \mathbf{X}_{l+1} \ast \mathbf{W}_l^T 
\label{eq:gradient_propagation}
\end{equation}

Here, \( \nabla \mathbf{X}_{l+1} \) is the gradient propagated back from layer \( l+1 \), while \( \mathbf{W}_l \) represents the weights of layer \( l \). The corresponding update to the weights is computed as:

\begin{equation}
\nabla \mathbf{W}_l = \hat{\mathbf{X}}_l^T \ast \nabla \mathbf{X}_{l+1} 
\label{eq:weight_update}
\end{equation}

Where \( \nabla \mathbf{W}_l \) represents the weight updates for layer \( l \), and \( \hat{\mathbf{X}}_l \) denotes the dequantized activations. This approach optimizes memory usage during training while preserving the precision required for gradient computations in both the forward and backward passes.

%% file: 3-Model.tex

\section{DAF Design}

\subsection{Overview}\label{sec:overview}
During training, activations represent the intermediate outputs of each layer, temporarily stored to compute weight gradients during backpropagation. Unlike weights and optimizer states, activations dominate memory usage, particularly on resource-constrained mobile devices. 
Our proposed DAF decouples the storage of activations from their computation. As illustrated in Figure~\ref{fig:overview} (a), activations are dynamically stored at varying bit widths, depending on their sensitivity. Before the backward computation, they are dequantized to the bit width required for gradient calculations, ensuring both memory efficiency and computational accuracy.

In this paper, we propose an end-to-end dynamic activation storage framework to reduce the memory footprint of on-device training. First, we introduce the dynamic activation control strategy and its corresponding system-level support in Section~\ref{sec:dynamic}. Next, we address the bit layout conversion problem by leveraging the collaborative capabilities of the CPU and GPU in Section~\ref{sec:bitconversion}. Finally, we tackle the issue of memory management by proposing a paging-based memory management method that leverages red-black tree indexing in Section~\ref{sec:mem}.

\subsection{Dynamic Quantization Control}\label{sec:dynamic}

To dynamically control quantization bit-width, existing research develops sensitivity or importance metrics calculated for each activation during every training iteration. These metrics guide dynamic bit-width adjustments, optimizing the total importance value while staying within memory or time budgets~\cite{actnn,gact,elastictrainer}.
Notably, all existing methods for computing quantization scale, as well as sensitivity or importance metrics, such as quantization error, quantization variance, and tensor magnitude, require time-intensive collective reduction operations.
These operations traverse all values in a tensor to compute results like the minimum, maximum, or magnitude. As memory-bound processes, they are slower than standard DNN computations and, in extreme cases, can take as long as the original computation itself. Since this procedure is repeated for every activation in every training iteration, it introduces substantial time overhead.

Collective reduction operations are inherently GPU-unfriendly due to the architecture of GPUs. To achieve data parallelism, GPUs distribute data across multiple threads, with each thread storing its data in private registers. While threads within a block can access shared memory, they cannot access the registers of other threads, and blocks cannot access each other's shared memory. This data isolation significantly hampers the efficiency of reduction operations, which require traversing and aggregating data across the entire tensor, limiting the GPU's ability to fully leverage parallelism.
To address this challenge, DAF employs a hybrid approach that combines parallel reduction with atomic reduction. This design enables efficient data exchange and accelerates collective reduction operations across different mobile and embedded platforms.

\begin{figure}[!t]
    \centering
    \includegraphics[width=0.9\linewidth]{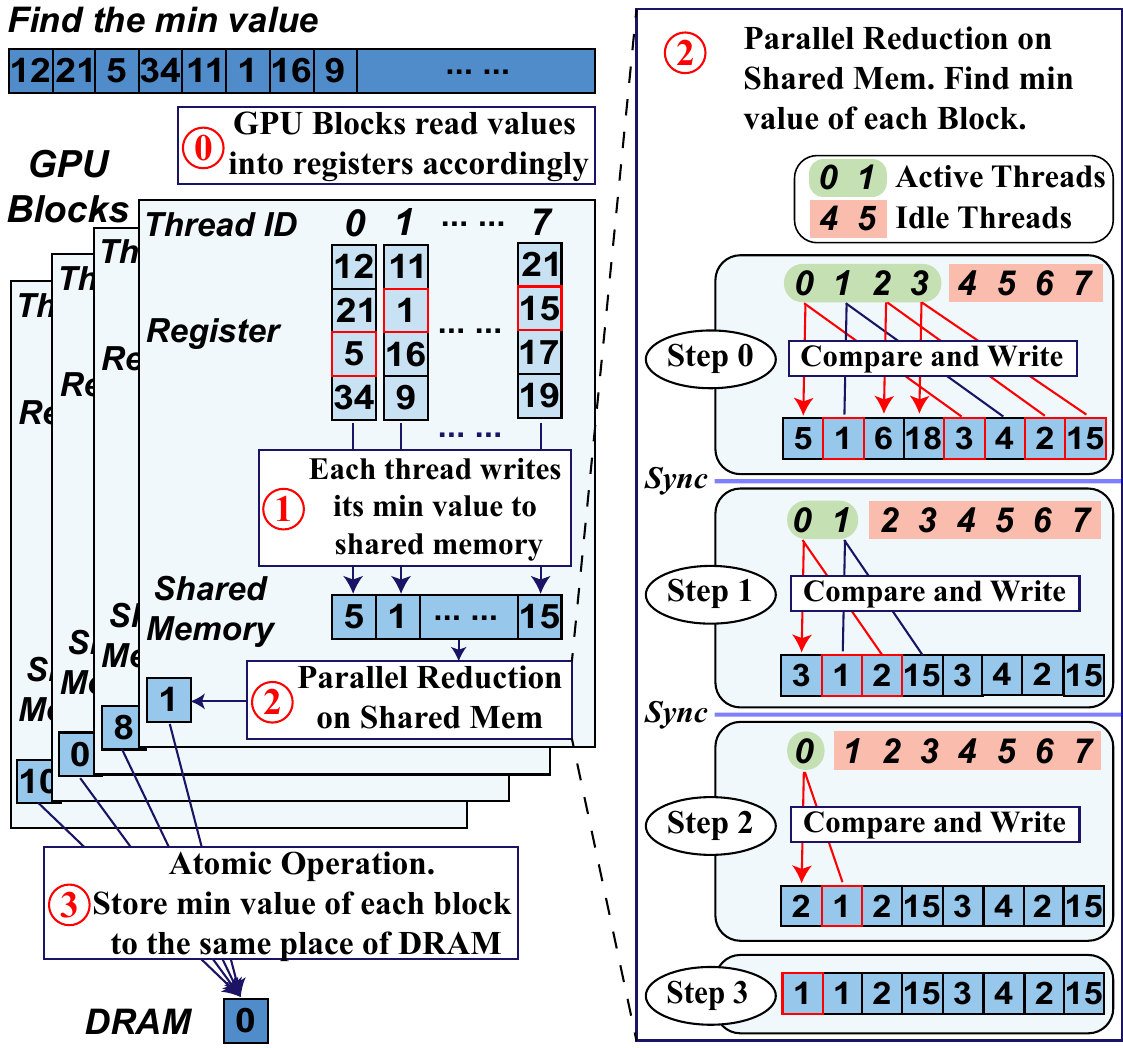}
    \vspace{-0.3cm}  
    \caption{\small Collective reduction operation example.}
    \label{fig:reduction}
\end{figure}

\subsubsection{Parallel Reduction Operation.}
The purpose of parallel reduction is to enable communication between threads or blocks by performing iterative computations at the lower level of the memory hierarchy. For example, to find the minimum value among all threads’ registers within a block (as shown in Figure~\ref{fig:reduction}, Stage 2), a tree-based reduction approach is employed.
In this method, threads collaboratively reduce shared memory arrays to a single minimum value per block. During each reduction step, the number of active threads is halved.
Each active thread compares the minimum values stored by two threads from the previous step and updates its own stored value accordingly. A synchronization step is performed at the end of each iteration to ensure all threads remain in sync. 
This process continues iteratively, halving the number of threads at each stage, until only one thread remains, holding the final minimum value for the block. 
This tree-based reduction efficiently reduces the problem size logarithmically, with a time complexity of $O(\log_2(N\_thread))$, where $N\_thread$ denotes the number of threads per block. 
To enable communication between blocks, the minimum value from each block is written back to DRAM. The parallel reduction process is then repeated across blocks to compute the global minimum value.

Parallel reduction offers several advantages, including the elimination of thread contention and efficient utilization of parallelism during computation. However, it also comes with notable limitations. For large tensors, parallel reduction must handle a substantial number of blocks, resulting in a large amount of intermediate data being written back to DRAM after each reduction pass. 
To further reduce the data, another parallel reduction kernel must be launched, and this process is repeated iteratively until only a single value remains as the final result.
While server GPUs, equipped with high-bandwidth global memory, handle this repeated DRAM read-and-write cycle relatively well, it becomes a major bottleneck on mobile and embedded SoCs. On these systems, all intermediate data must be written back to DRAM and reloaded for each iteration, leading to considerable time overhead due to the lower DRAM bandwidth.
For example, the NVIDIA 3090 with GDDR6X memory offers a bandwidth of 936 GB/s, whereas the iQOO Neo3, using LPDDR4X memory, achieves only 51.2 GB/s.
Moreover, mobile devices face bandwidth constraints across all memory levels. 
During the parallel reduction process in shared memory, each step requires thread synchronization. These frequent synchronizations introduce considerable overhead, making the process particularly time-consuming on mobile platforms.

To address this issue, in addition to the parallel reduction operation, DAF also provides an alternative option using atomic operations.

\SetKwComment{Comment}{/* }{ */}
\begin{algorithm}[t]
\vspace{-0.1cm}
\caption{\small Atomic Minimum for Floats}\label{alg:atomic_min_float}
\footnotesize
\textbf{Input}: Memory address \texttt{address}, value \texttt{val} \;
\textbf{Output}: Updated float value at \texttt{address} \;
\BlankLine
\texttt{address\_as\_uint} $\gets$ reinterpret \texttt{address} as unsigned int pointer \;
\texttt{old} $\gets$ \texttt{*address\_as\_uint} \;
\Repeat{\texttt{assumed} equals \texttt{old}}{
    \texttt{assumed} $\gets$ \texttt{old} \;
    \texttt{new\_val} $\gets$ \texttt{fminf(\_\_uint\_as\_float(old), val)} \;
    \texttt{old} $\gets$ \texttt{atomicCAS(address\_as\_uint, assumed, \_\_float\_as\_uint(new\_val))} \;
}
\Return \texttt{\_\_uint\_as\_float(old)} \;
\end{algorithm}

\subsubsection{Atomic Reduction Operation.}
The Atomic Reduction Operation also facilitates communication between threads and blocks while ensuring reduction values are stored safely and efficiently at lower levels of the memory hierarchy.
Like traditional atomic operations, it allows multiple threads to safely access the same memory address without data races. As shown in Stage 3 of Figure~\ref{fig:reduction}, the minimum values from multiple blocks are directly compared and written to a single shared address in DRAM. 
This approach contrasts with parallel reduction, where minimum values are first written to separate DRAM addresses and then reloaded for subsequent reduction steps.

While there is no native atomic operation for compare-and-swap with float values, we can implement this functionality using integer-based atomic operations and bitwise manipulation.
As shown in Algorithm~\ref{alg:atomic_min_float}, the float value at a given memory address is first reinterpreted as an integer using type casting. This allows us to leverage the atomic compare-and-swap operation ($atomicCAS$ for CUDA and $atomic\_cmpxchg$ for OpenCL), which operates natively on integers. By treating the bit representation of the float as an integer, atomic updates can be performed safely and effectively. The $atomicCAS$ operation compares the current value at the memory address (interpreted as an integer) with the new value, and conditionally updates it if the new value is smaller. After the comparison and potential update, the integer is reinterpreted back into a float, preserving the original bitwise structure of the floating-point value.


However, the atomic reduction operation may introduce contention, which highlights the trade-offs between atomic and parallel reduction methods. 
By tailoring reduction strategies to the specific characteristics of each device, these two methods can be combined at different levels (block/thread) to minimize overall time overhead.
For example, the Jetson AGX Xavier has only 8 Streaming Multiprocessors (SMs), whereas the NVIDIA 3090 features 82 SMs. This substantial difference in hardware resources means that, even with an identical number of blocks in a computation kernel, far fewer blocks run concurrently on the Jetson AGX Xavier than on the NVIDIA 3090.
As a result, the Jetson Xavier exhibits much lower contention for atomic reduction operations at the block level. In contrast, server GPUs like the NVIDIA 3090 are prone to contention due to the larger number of concurrent blocks competing for the same atomic memory addresses. This difference in contention sensitivity makes block-level atomic reduction operations more practical and efficient on embedded GPUs, such as the Jetson Xavier, while being less effective on server GPUs.

A hybrid reduction operation, as illustrated in Figure~\ref{fig:reduction}, achieves low overhead on devices such as the Jetson AGX Xavier and Jetson AGX Orin.
However, on mobile GPUs, where synchronization efficiency is significantly lower, using atomic operations first at the thread level and then at the block level outperforms the hybrid approach in approximately 15\% of cases. This finding underscores the effectiveness of atomic operations in specific scenarios, particularly when synchronization overhead is a critical concern. In practice, profiling can be performed offline to determine and configure the optimal reduction strategy for each operation.

\subsubsection{Kernel Fusion.}
To achieve further speedup, we categorize the computations involved in dynamic quantization control into two types: element-wise operations and reduction operations.
\textit{Element-wise operations}, as shown in Equation~\eqref{eq:quant}, compute each output element using only a single corresponding input element from the tensor. These operations are highly parallelizable and can be efficiently fused with forward and backward computations.

For \textit{reduction operations}, we optimize computation by performing the reduction for \( X_{l+1} \), the output of layer \( l \) during the forward pass. As each thread computes the corresponding value of \( X_{l+1} \), the reduction result is simultaneously calculated using the hybrid reduction operation discussed earlier. This strategy minimizes the need to reload values from DRAM, significantly reducing the time overhead associated with reduction operations.

\SetKwComment{Comment}{/* }{ */}
\begin{algorithm}[t]
\caption{\small Unpacking \texttt{uint4} to \texttt{uint32} with SIMD}\label{algo:unpack_simd}
\footnotesize
\textbf{Input}: \texttt{input} - an array of \texttt{uint32}, where each \texttt{uint32} contains eight packed \texttt{uint4} values \;
\textbf{Output}: \texttt{output} - an array of \texttt{uint32}, where each element is a \texttt{uint4} unpacked from the input \;
\BlankLine
\For{\texttt{i} $\gets 0$ \KwTo \texttt{num\_uint32} in steps of 4}{
    \Comment{Load four \texttt{uint32} values into NEON registers}
    \texttt{input\_vec} $\gets$ \texttt{vld1q\_u32(\&input[i])} \;
    \For{\texttt{j} $\gets 0$ \KwTo 7}{
        \Comment{Shift left by $j \times 4$ bits and then right by 28 bits to isolate the \texttt{uint4}}
        \texttt{shifted} $\gets$ \texttt{vshrq\_n\_u32(vshlq\_n\_u32(input\_vec, j * 4), 28)} \;
        \Comment{Store the extracted \texttt{uint4} in the output array}
        \texttt{vst1q\_u32(\&output[(i * 8) + (j * 4)], shifted)} \;
    }
}
\Return \texttt{output} \;
\end{algorithm}

\subsection{Bit Layout Conversion}\label{sec:bitconversion}

In the previous subsection, we discussed the support for collective reduction operations, which are essential for accelerating quantization and computing the associated dynamic control metrics.
However, reducing memory usage effectively requires storing the quantized activations in their corresponding bit-width format and concatenating them. 
For example, if the selected quantization bit width is 4, the previous steps computed the values quantized to 4 bits, but these values are still stored in \texttt{float32} format. To optimize storage, we must first convert these values from \texttt{float32} to uint representation, then pack every eight \texttt{uint4} values into a single \texttt{uint32}. This process involves two key steps: bit conversion and layout concatenation. However, achieving this memory reduction comes with significant time overhead.

To address this, DAF introduces a layout transformation library with two configurable options tailored to different hardware and DNN operators. The first option leverages fully fused and optimized GPU kernels, while the second leverages the SoC's unified memory for collaborative packing and unpacking between the CPU and GPU. A one-time profiling process determines the most efficient option for minimizing overhead on a given system.

\subsubsection{Bit Conversion.}
On some mobile and embedded devices, the instruction throughput for bit conversion can be considerably lower compared to floating-point operations~\cite{nvidia_instructions,arm_instructions}.
Therefore, DAF offers an option to transform bit conversion into computation, effectively reducing the time overhead. The \texttt{float32} data storage format consists of 32 bits organized into three components: 1 bit for the sign, 8 bits for the exponent (stored in a biased format with a bias of 127), and 23 bits for the mantissa, which represents the significant digits of the number. The mantissa is normalized with an implicit leading 1, enabling a precise representation of a wide range of real numbers:
\begin{equation}
\text{value} = (-1)^{\text{sign}} \ast 2^{(\text{exponent} - 127)} \ast \left(1 + \frac{\text{mantissa}}{2^{23}}\right)
\label{eq:float32_representation}
\end{equation}


It is straightforward to see that for a \texttt{uint32} number represented as \texttt{0bX}, with a quantization bit width of 4, the process involves concatenating 19 zero bits to these 4 bits to form the mantissa. Next, 8 bits with the value \texttt{0b10010110} (decimal 150) are appended as the exponent, followed by a 0 as the sign bit. This operation can be simplified to a bitwise OR between the \texttt{uint32} data and the constant \texttt{0x4B000000}.
The resulting \texttt{float32} value is given by:
\begin{equation}
\text{value} = (-1)^{0} \ast 2^{(150 - 127)} \ast \left(1 + \frac{X}{2^{23}}\right) = 2^{23} + X  
\label{eq:uint4_to_float32}
\end{equation}
In other words, by prefixing any \texttt{uint4} with a fixed 28-bit sequence, the \texttt{uint32} value $X$ can be converted into its \texttt{float32} representation as $X + 2^{23}$, Once this concatenation is performed, subtracting $2^{23}$ directly yields the \texttt{float32} representation of $X$.
Converting \texttt{float32} back to \texttt{uint32} is a straightforward reverse process. 
These bit conversion calculations are performed element-wise, making them well-suited for fusion into GPU computation kernels for parallel processing with minimal overhead.

\subsubsection{Activation Package.}
To efficiently store data, \texttt{uint4} values are extracted from \texttt{uint32} and packed. DAF can leverage the SoC's unified memory and latency-hiding capabilities between the CPU and GPU, enabling the CPU to process \texttt{uint32} values generated by the GPU with minimal overhead. Our implementation utilizes a CPU-based approach with SIMD, specifically ARM's NEON instruction set, for efficient extraction and packing operations. By integrating SIMD's parallelism with OpenMP-based multithreading, this method achieves optimal CPU resource utilization.
The example demonstrates unpacking 8 \texttt{uint4} values from a packed \texttt{uint32} into 8 separate \texttt{uint32} values, while the reverse operation, packing 8 \texttt{uint32} values into a single \texttt{uint32}, is simpler.
As described in Algorithm \ref{algo:unpack_simd}, the extraction process involves loading four \texttt{uint32} values into NEON registers and using a sequence of bit shifts to isolate the \texttt{uint4} values. Each \texttt{uint32} contains 8 packed \texttt{uint4} values, each occupying 4 bits. To extract a specific \texttt{uint4}, we first left-shift the \texttt{uint32} by $j \times 4$ bits (where $j$ is the position of the target \texttt{uint4}), moving the desired \texttt{uint4} to the most significant bits. A subsequent right shift by 28 bits isolates the \texttt{uint4} in the least significant 4 bits. Repeating this operation for all positions $j$ extracts all 8 \texttt{uint4} values, which are then stored as full 32-bit \texttt{uint32} values in the output array.


\subsubsection{Unified CPU-GPU Layout.}

\begin{figure}[!t]
    \centering
    \includegraphics[width=0.95\linewidth]{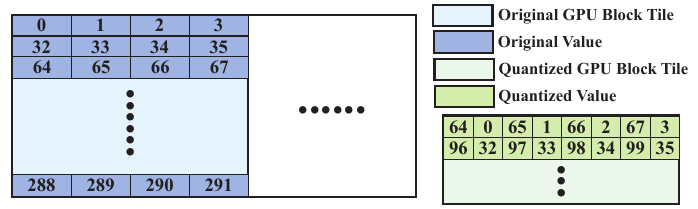}
    \caption{\small Unified Bit Layout for efficient CPU packing and GPU unpacking.}
    \label{fig:unified_layout}
\end{figure}

To further reduce memory consumption without incurring significant time overhead, our system packs data using the CPU during the forward pass and unpacks it using the GPU during the backward pass. This strategy avoids the need to explicitly store unpacked activations in DRAM. However, this approach creates a bit layout mismatch between the CPU and GPU as shown in Figure~\ref{fig:motivation_3_keys} (b), where the CPU layout consists of an interleaved alternating sequence.


The challenges are threefold. First, online layout transformation is impractical due to its high time overhead. Second, a SIMD-based CPU packing algorithm inherently produces an interleaved order, whereas avoiding SIMD instructions to achieve a non-interleaved layout would drastically increase CPU packing time, creating a performance bottleneck. Third, directly utilizing the interleaved layout generated by the CPU for unpacking and gradient computation on the GPU is problematic, as the interleaved format can disrupt GPU computation tiles, leading to inefficient global memory access patterns and degraded performance.

To address these challenges, DAF introduces a unified layout for bit (un)packing, as illustrated in Figure~\ref{fig:unified_layout}. This design leverages GPU computation-tile information to create a coarse-grained layout that optimizes memory access patterns during GPU computations. Simultaneously, within each data tile, the fine-grained layout incorporates the interleaved order produced by the SIMD-based CPU packing algorithm, ensuring its compatibility and maintaining packing efficiency.

\subsection{Memory Management}\label{sec:mem}
In this section, we discuss the challenges of memory management in resource-constrained environments. In static graph frameworks like TensorFlow, memory reuse and optimization techniques are commonly employed to reduce peak memory usage~\cite{tensorflow}. However, in dynamic activation quantization training, the frequent computation, storage, and deletion of activations not only exacerbate memory fragmentation but also make it challenging for TensorFlow’s static analysis to preallocate suitable memory for each operation. 
Some dynamic activation quantization studies attempt to optimize memory usage by manually clearing caches during training~\cite{actnn,gact}. However, these approaches often yield suboptimal results due to the inability to effectively address memory fragmentation. 
To address these issues, we propose a page-like memory management scheme that efficiently handles memory allocation as well as the dynamic storage and removal of activations.

\subsubsection{Dynamic Activation Storage Control.}\label{sec:greedy}


\begin{figure}[!t]
    \centering
    \includegraphics[width=0.8\linewidth]{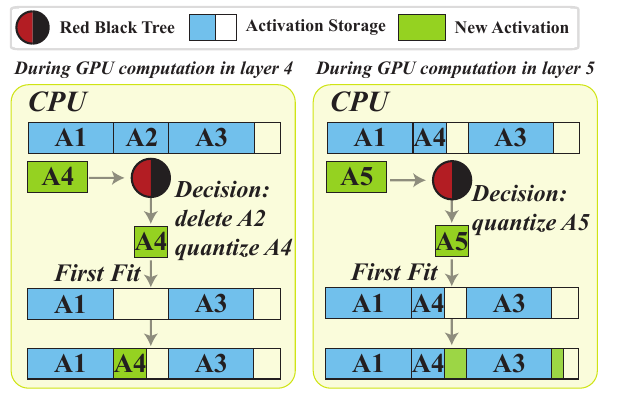}
    \caption{\small Memory management example. Dynamic quantization storage managed by Red-Black Tree w/ First-Fit page allocation.}
    \label{fig:page}
\end{figure}

We begin by explaining DAF's Dynamic Activation Storage Control strategy. As discussed in Sec.~\ref{sec:dynamic}, DAF calculates the sensitivity or importance value of each activation during its computation in the forward propagation phase. This value serves as a critical criterion for determining the bit width of the current activation. A bit width of zero indicates that the activation will not be stored.
Broadly speaking, dynamic activation quantization algorithms can be unified under the objective of maximizing the total importance of all stored activations within given time and memory budgets~\cite{actnn,gact,elastictrainer}. This can be formulated as a bi-objective knapsack problem, where the goal is to identify the accuracy Pareto front under time and memory constraints. As a classic combinatorial optimization problem~\cite{multiobjective}, DAF employs a greedy algorithm for acceleration.
Specifically, when selecting activations, DAF follows these steps: first, it determines which constraint, memory or time, is closer to being exhausted by comparing their remaining proportions.  
Next, activations are sorted based on the tightest constraint. For example, if the memory constraint is more critical, activations are prioritized by the importance/memory\_footprint.
Finally, if multiple candidates satisfy the primary constraint, the secondary constraint is considered for further prioritization.  
This greedy algorithm enables DAF to efficiently compute bit widths and manage the addition or removal of activations.


\begin{figure}[!t]
\centering
{\includegraphics[width=0.98\linewidth]{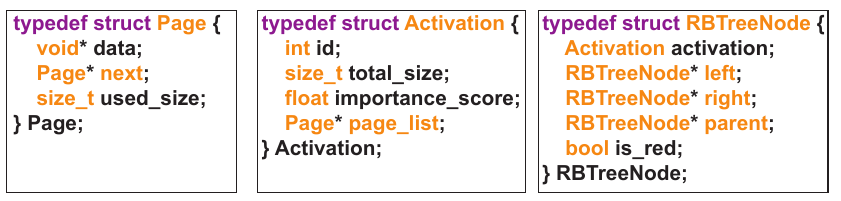}}
\vspace{-0.5cm}
\caption{\small Red-Black Tree API example.
}
\label{fig:api}
\end{figure}

\subsubsection{Page Allocation.}\label{pageallocation}
To optimize memory usage, DAF is inspired by TensorFlow’s static memory design and incorporates offline analysis to evaluate the lifecycle and reuse patterns of non-activation memory. For activations, DAF allocates a contiguous memory block based on the memory budget. During runtime, activations are managed using a page-like indexing approach, enabling efficient storage and retrieval within the allocated memory block.

This approach addresses the limitation of TensorFlow, which cannot dynamically analyze the size of each activation and thus fails to preallocate memory efficiently. By introducing a predefined memory budget and manually managing the storage of activations, DAF maximizes memory utilization. Furthermore, the use of a page-based allocation and deallocation scheme helps to mitigate fragmentation within the activation storage, as illustrated in Figure~\ref{fig:page}.

\subsubsection{Page Table Maintenance.}\label{redblacktree}
Unlike traditional pages, DAF’s pages are augmented with an importance attribute, which corresponds to the value computed in the dynamic quantization control module in Sec.~\ref{sec:dynamic}. When storing an activation, DAF dynamically determines its bit width based on its importance and evaluates whether to evict existing activations to make room for the new one. To enable efficient management of these operations, DAF employs a red-black tree to maintain the page table. As shown in the API representation, the red-black tree’s key is a custom activation structure, which includes attributes such as the activation ID, current size, importance score, and the location within the linked list of pages.

Under both time and memory budgets, two red-black trees must be maintained. When one tree is updated, the other must also be updated to ensure consistency. Combining the algorithm described in Sec.~\ref{sec:greedy}, the page management process is shown as Algorithm~\ref{alg:page_management}.

\begin{algorithm}[t]
\small
\caption{\small Page Management with Dual Red-Black Trees}
\label{alg:page_management}
\KwIn{Activations \(\{A_i\}\), moving average importance \(\{\text{MA}_i\}\), budgets \(B_\text{mem}\), \(B_\text{time}\)}
\KwOut{Updated page table and red-black trees}
\SetKwBlock{Initialize}{Initialization}{end}
\SetKwBlock{Estimate}{Pre-iteration Estimation}{end}
\SetKwBlock{Forward}{Forward Pass}{end}


\Estimate{
    \For{\(A_i \in \{A_i\}\)}{
        \While{\(B_\text{mem} < \text{UsedMem} \lor B_\text{time} < \text{UsedTime}\)}{
            \(A_{\text{max}} \gets \texttt{SelectMax}(\texttt{TighterTree})\)\;
            \eIf{ \(\texttt{TighterTree} == \texttt{MemTree}\)}{
                \(\text{UpdateSize}(A_{\text{max}})\)\;
            }{
                \(\text{Evict}(A_{\text{max}})\)\;
            }
        }
        \(\texttt{MemTree}.\text{Insert}(A_i, \text{key}=\frac{\text{MA}_i}{\text{size}_i})\)\;
        \(\texttt{TimeTree}.\text{Insert}(A_i, \text{key}=\frac{\text{MA}_i}{\text{time}_i})\)\;
    }
}

\Forward{
    \While{\text{Receiving activation } \( (A_i, \text{importance}_i)\)}{
        \(\texttt{MemTree}.\text{UpdateKey}(A_i, \frac{\text{importance}_i}{\text{size}_i})\)\;
        \(\texttt{TimeTree}.\text{UpdateKey}(A_i, \frac{\text{importance}_i}{\text{time}_i})\)\;

        \While{\(B_\text{mem} < \text{UsedMem} \lor B_\text{time} < \text{UsedTime}\)}{
            \(A_{\text{max}} \gets \texttt{SelectMax}(\texttt{TighterTree})\)\;
            \eIf{\(A_{\text{max}} \in \texttt{Unallocated}\) and \(\texttt{TighterTree} == \texttt{MemTree}\)}{
                \(\text{UpdateSize}(A_{\text{max}})\)\;
            }{
                \(\text{Evict}(A_{\text{max}})\)\;
            }
        }

        \(\text{location} \gets \texttt{FirstFit}(\texttt{FreePages}, A_i)\)\;
        \(\texttt{InsertPage}(\texttt{MemTree}, \texttt{TimeTree}, A_i, \text{location})\)\;
    }
}
\end{algorithm}

\subsubsection{Dynamic Memory Budget.}\label{sec:dynamic_mem}

On edge and mobile devices, where GPUs and CPUs share unified memory, the GPU memory budget cannot always remain fixed. DAF is designed to support dynamically varying memory budgets, enabling seamless adjustments to accommodate changes in available memory. Inspired by the design of C++ vector containers, DAF adopts an incremental memory allocation strategy for activation storage under dynamic memory budget scenarios. 
For instance, if the memory budget fluctuates between 500 MB and 800 MB, DAF initially allocates 100 MB for activation storage. When this allocation is fully utilized, additional 100 MB increments are requested as needed, up to the upper limit of the budget. Conversely, when the memory budget decreases, DAF scales down the allocated activation storage in decrements of 100 MB until it fits within the new budget constraints. 

Whenever the memory budget changes, the budget parameter in Algorithm~\ref{alg:page_management} is updated accordingly. DAF immediately maintains the red-black tree and performs any necessary memory copying to ensure the current used memory remains within the updated budget. By selecting an appropriate step size for memory allocation adjustments, the overhead introduced by these operations is minimal and can be effectively masked by the GPU's forward computations. Our experiments indicate that a step size of 100 MB is a reasonable and practical choice.

%% file: 4-Evaluation.tex
\begin{table*}[ht]
\centering
\footnotesize 
\caption{\small Comparison of accuracy and memory usage across methods and datasets on Jetson AGX Xavier.}
\resizebox{\textwidth}{!}{
\begin{tabular}{|c|cc|cc|cc|cc|cc|cc|cc|cc|}
\hline
 & \multicolumn{6}{c|}{\textbf{ResNet-18 (Train-from-scatch)}} & \multicolumn{6}{c|}{\textbf{RoBERTa-base (LoRA fine-tuning)}} & \multicolumn{4}{c|}{\textbf{GPT2-M (LoRA fine-tuning)}} \\ \hline
 & \multicolumn{2}{c|}{CIFAR-10~\cite{CIFAR-10}} & \multicolumn{2}{c|}{CIFAR-100~\cite{CIFAR-100}} & \multicolumn{2}{c|}{ImageNet1K~\cite{ImageNet1K}} & \multicolumn{2}{c|}{STSB~\cite{STSB}} & \multicolumn{2}{c|}{MRPC~\cite{MRPC}} & \multicolumn{2}{c|}{SST2~\cite{SST2}} & \multicolumn{2}{c|}{E2E~\cite{E2E}} & \multicolumn{2}{c|}{WebNLG~\cite{WebNLG}} \\ \hline
 & \textbf{acc (\%)} & \textbf{mem (MB)} & \textbf{acc (\%)} & \textbf{mem (MB)} & \textbf{acc (\%)} & \textbf{mem (MB)} & \textbf{acc (\%)} & \textbf{mem (MB)} & \textbf{acc (\%)} & \textbf{mem (MB)} & \textbf{acc (\%)} & \textbf{mem (MB)} & \textbf{acc (\%)} & \textbf{mem (MB)} & \textbf{acc (\%)} & \textbf{mem (MB)} \\ \hline
Original & 92.7 & 1196 & 75.6 & 917 & 70.1 & 5110 & 90.7 & 3470 & 89.2 & 3472 & 93.1 & 3470 & 69.7 & 8385 & 58.1 & 8163 \\ \hline
Fixed 8 Bits & 92.4 & 670 & 75.1 & 457 & 69.5 & 2715 & 90.1 & 2390 & 88.5 & 2390 & 92.7 & 2391 & 69 & 5928 & 56.8 & 5915 \\ \hline
ElasticTrainer & 92.7 & 1210 & 75.6 & 917 & 70.1 & 5112 & 90.7 & 3470 & 89.2 & 3472 & 93.1 & 3473 & 69.7 & 8412 & 58.1 & 8206 \\ \hline
ActNN & 91.7 & 900 & 74.5 & 854 & 69.1 & 3905 & 89.5 & 1912 & 88.2 & 2321 & 92.1 & 1968 & 68.8 & 5620 & 57.1 & 6402 \\ \hline
\textbf{DAF} & 91.7 & \textbf{52} (\textbf{22.9\(\times\)}) & 74.7 & \textbf{49} (\textbf{18.7\(\times\)}) & 69.1 & \textbf{442} (\textbf{11.6\(\times\)}) & 89.7 & \textbf{188} (\textbf{18.4\(\times\)}) & 88.3 & \textbf{310} (\textbf{11.2\(\times\)}) & 92.2 & \textbf{289} (\textbf{12.0\(\times\)}) & 68.9 & \textbf{417} (\textbf{20.1\(\times\)}) & 57.2 & \textbf{434} (\textbf{18.8\(\times\)})\\ \hline
\end{tabular}
}
\label{tab:end2end-acc-mem}
\end{table*}

\begin{figure*}[!htbp]
\centering
{\includegraphics[width=0.98\linewidth]{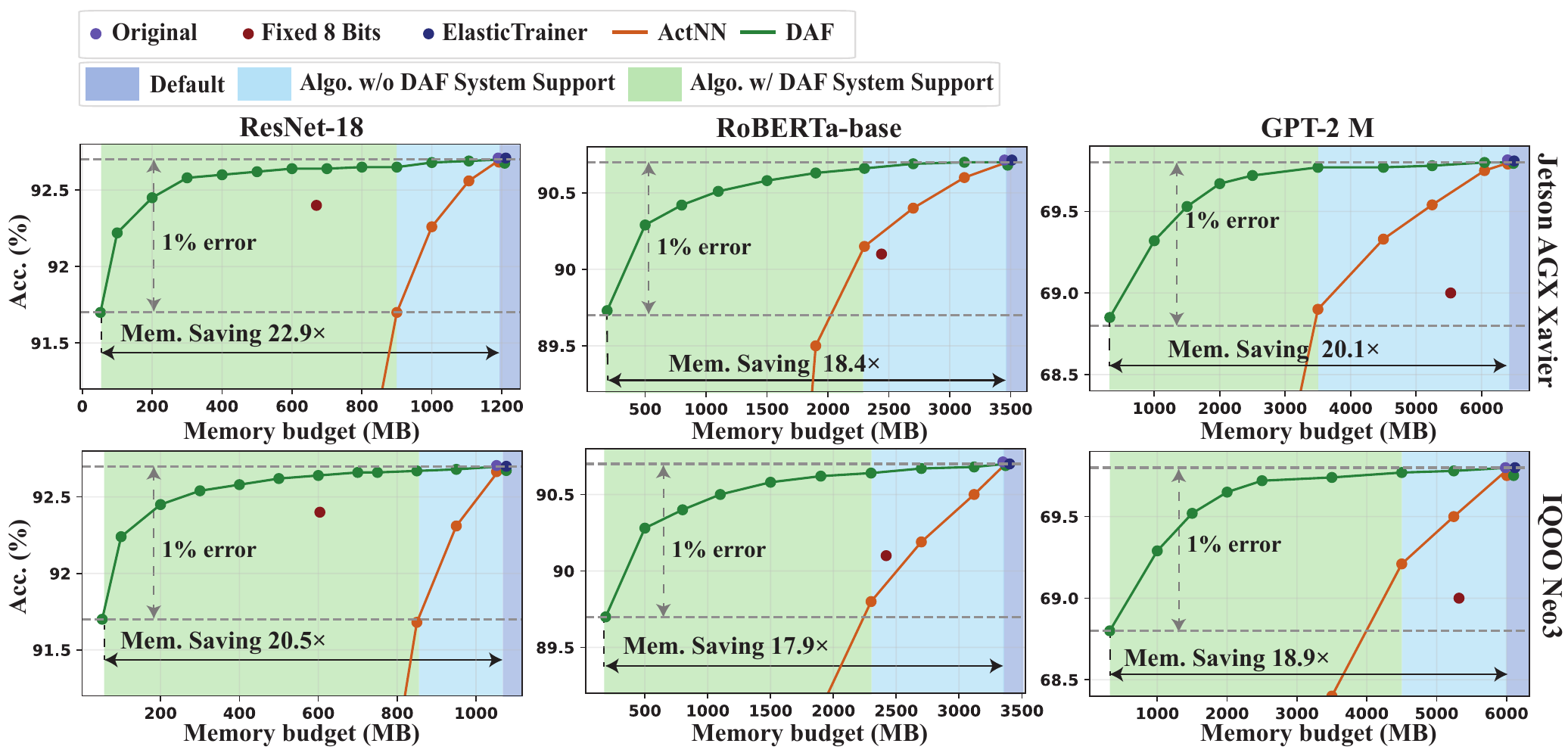}}
\caption{\small Comparison of accuracy under varying memory budgets.}
\label{fig:end2end-acc-mem}
\end{figure*}

\begin{figure}[!t]
\centering
{\includegraphics[width=0.95\linewidth]{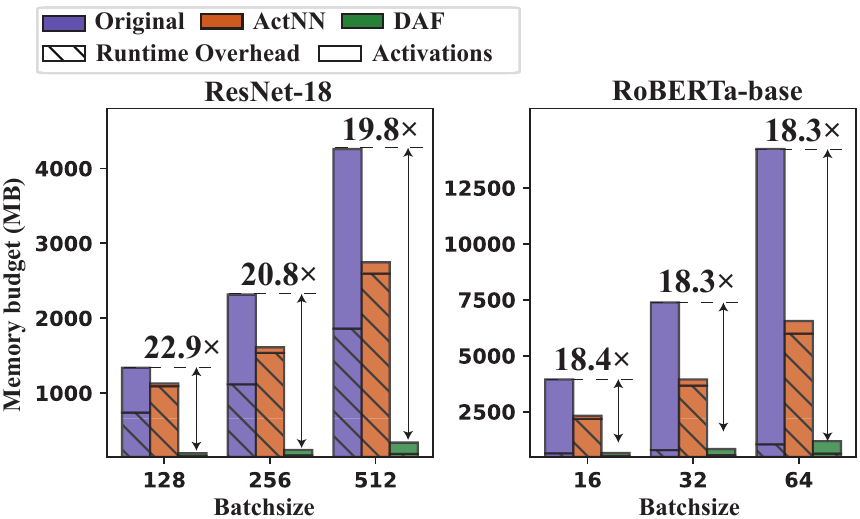}}
\vspace{-0.35cm}
\caption{\small Memory budget with varying batch sizes for ResNet-18 and RoBERTa-base using Jetson AGX Xavier, under an accuracy drop of <1\%.
}
\label{fig:mem-batch}
\end{figure}

\begin{figure*}[!t]
\centering
{\includegraphics[width=0.98\linewidth]{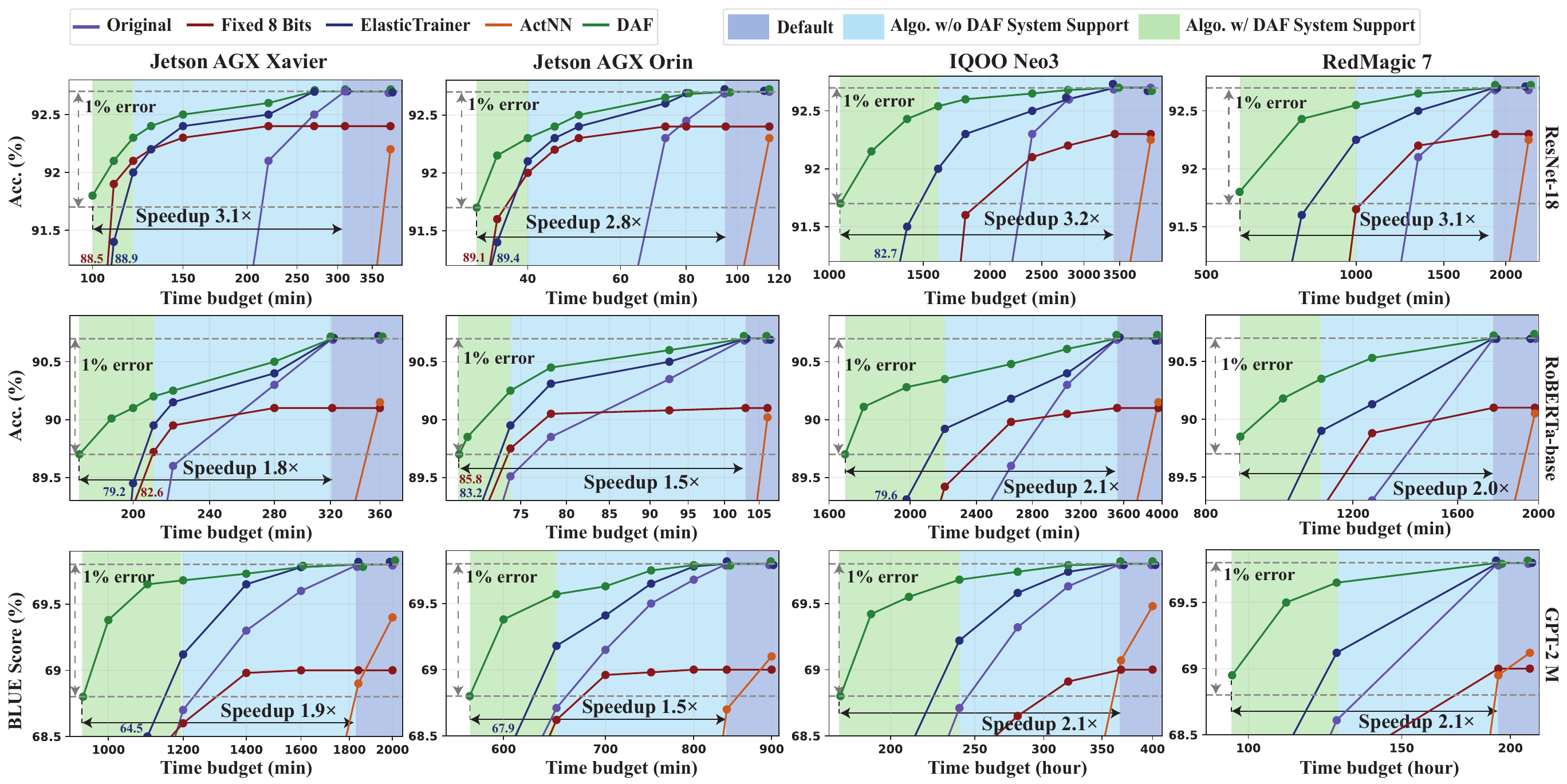}}
\caption{\small Comparison of accuracy under varying time budgets.
}
\label{fig:end2end-acc-time}
\end{figure*}

\vspace{-0.5cm}
\section{Evaluation}\label{sec:evaluation}
\subsection{Experimental Setup}
\noindent {\bf Models and Devices.} 
We evaluate DAF on three deep learning models: one convolution-based model (ResNet-18~\cite{resnet}) and two transformer-based models (RoBERTa-base~\cite{roberta} and GPT-2 Medium~\cite{gpt2}). For ResNet-18, we use a train-from-scratch approach, while for RoBERTa-base and GPT-2 Medium, we adopt LoRA~\cite{lora} fine-tuning. 
The evaluations are conducted across four platforms: two embedded devices, the Jetson AGX Orin with 32GB memory and the Jetson AGX Xavier with 16GB memory, and two mobile devices, the iQOO Neo 3 with a Snapdragon 865 chipset and 12GB memory and the RedMagic 7 with a Snapdragon 8 Gen 1 chipset and 12GB memory. These experiments demonstrate the effectiveness of DAF across diverse network architectures, training methods, and hardware configurations.

\noindent {\bf Baselines.} We compare DAF against four baselines: Original, Fixed 8 Bits~\cite{int8train}, ElasticTrainer~\cite{elastictrainer}, and ActNN~\cite{actnn}. These represent different training approaches. The Original baseline corresponds to standard training, implemented using PyTorch on the Jetson platforms and OpenCL on the mobile platforms. Fixed 8 Bits stores both weights and activations as 8-bit without dequantizing them during computation.  
ElasticTrainer is a dynamic activation framework that skips computations for less important layers to reduce training time, effectively setting the activation bit width of these layers to zero. ActNN is another dynamic activation compression framework, designed to minimize activation storage. 
On Jetson platforms, we use baselines' open-source implementations. On mobile platforms, we implement their functionality with OpenCL and TVM.



\noindent {\bf Metrics.} 
In our experiments, we evaluate the performance of each method under the constraint of negligible accuracy drop (< 1\%). Specifically, we compare the minimum training time and training memory footprint achieved by each method to the default training time and memory footprint. 
Here, runtime memory footprint refers to the memory usage generated during the training phase, including both stored activations and the memory allocated but not actively utilized, such as fragmentation.
\subsection{End-to-end Results}
\noindent {\bf Accuracy and Memory Usage.} As shown in Table~\ref{tab:end2end-acc-mem}, across all networks, datasets, and training approaches, DAF consistently outperforms other methods in optimizing training memory usage while maintaining negligible accuracy loss. Compared to the original standard training method, our approach achieves memory savings ranging from 11.2\(\times\) to 22.9\(\times\), with the largest improvement observed in training ResNet-18 on CIFAR-10. This significant gain is attributed to CIFAR-10 being a relatively simple dataset, where the small resolution allows our quantization technique to achieve a higher level of compression. ElasticTrainer’s memory footprint is identical to that of the original method because it does not include any modules for actual memory optimization. Instead, ElasticTrainer focuses solely on skipping computations to accelerate training without addressing the practical memory-saving issue. Consequently, in scenarios where the evaluation prioritizes memory optimization without time constraints, ElasticTrainer defaults to the same training time and accuracy as the original method. This is reflected in this table, where ElasticTrainer shows no improvements in memory usage compared to the original baseline.


\begin{figure}[!t]
\centering
{\includegraphics[width=0.98\linewidth]{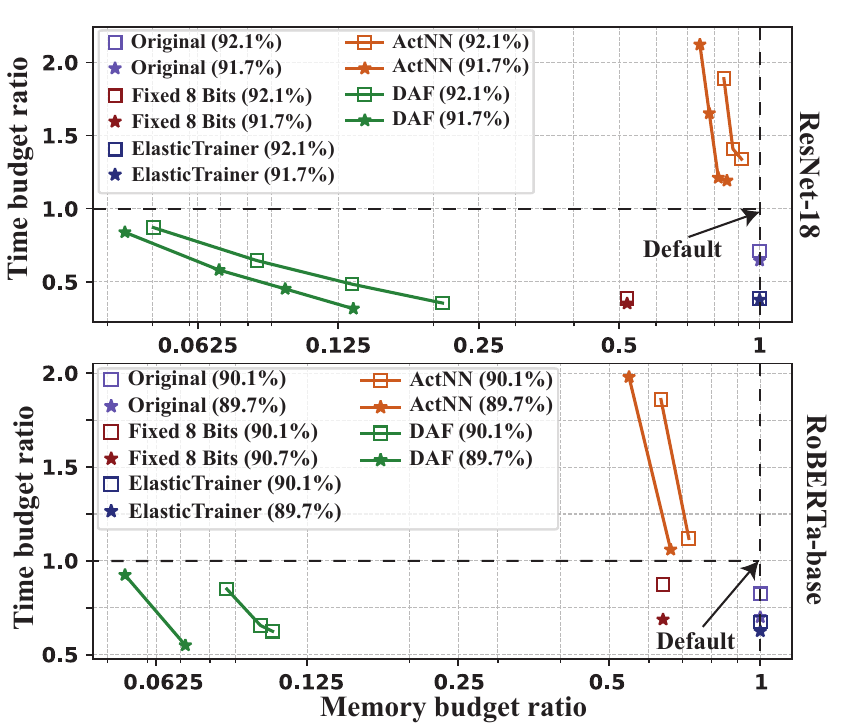}}
\vspace{-0.2cm}
\caption{\small Comparison of time and memory usage for various methods under different levels of accuracy drop, relative to the default setup, on Jetson AGX Xavier.}
\label{fig:end2end-time-mem}
\end{figure}

\noindent {\bf Accuracy under Varying Memory Budgets.} As shown in Figure~\ref{fig:end2end-acc-mem}, under scenarios where only the memory budget is constrained, DAF achieves memory savings ranging from \(17.9\times\) to \(22.9\times\) compared to the default training memory usage. The light blue sections in the figure represent the best possible memory savings achievable without DAF support, corresponding to the capabilities of ActNN. It is evident that DAF provides a significant advantage in memory savings over ActNN.
Figure~\ref{fig:mem-batch} illustrates the impact of varying batch sizes on the memory budget under an accuracy drop of <1\%. As shown, DAF consistently demonstrates stable and substantial memory reduction across diverse batch-size configurations.

\noindent {\bf Accuracy under Varying Time Budgets.} 
Figure~\ref{fig:end2end-acc-time} illustrates the accuracy trends when training time is constrained without limiting the memory budget. As shown, DAF achieves a significant speed-up compared to the Default time, which refers to the maximum time required by the original method to complete training. DAF demonstrates speed-ups ranging from \(1.5\times\) to \(3.2\times\) across various platforms.
Furthermore, compared to Fixed 8 Bits and ElasticTrainer, DAF achieves higher accuracy under the same training time constraints. 
In the figure, the blue and red numbers represent the accuracy of Fixed 8 Bits and ElasticTrainer, respectively, within the time budget required for DAF to reach a $1\%$ accuracy drop.
It is evident that both Fixed 8 Bits and ElasticTrainer exhibit significantly accuracy drops, underscoring the superiority of DAF.


\noindent {\bf Accuracy under Simultaneously Varying Memory-Time Budgets.}
Figure~\ref{fig:end2end-time-mem} illustrates the memory and time budgets required to achieve a given accuracy. Under a fixed memory budget, the time budget represents the total training time. Lower computation overhead allows more epochs to be trained within the same time.
Under simultaneous time and memory budget constraints, DAF achieves significantly lower memory usage compared to other methods while maintaining competitive training times at the same accuracy drop. In contrast, ActNN, lacking system support for memory management and collective-reduce acceleration, demonstrates a poor trade-off between memory and time. Its theoretical memory savings fail to effectively translate into training speedup. The manual cache-clearing strategy employed by ActNN proves ineffective in real-world scenarios.

\begin{figure}[!t]
\centering
\vspace{-0.1cm}
{\includegraphics[width=0.98\linewidth]{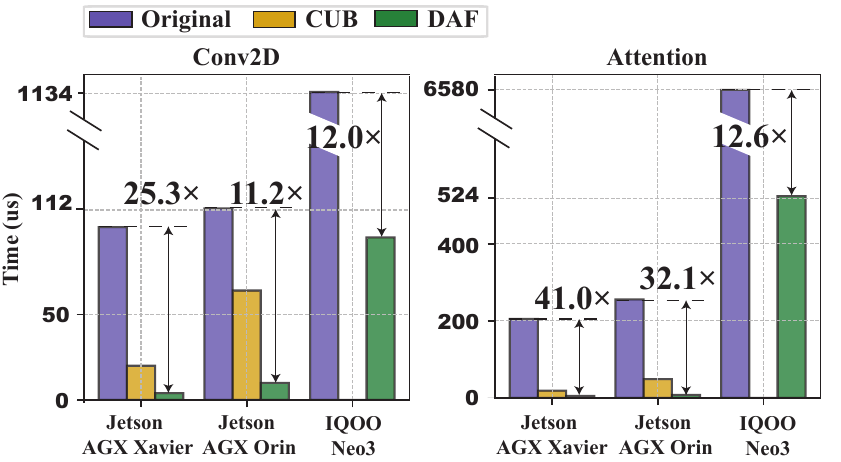}}
\vspace{-0.35cm}
\caption{\small Reduction overhead comparison on Conv2D and attention layer with different devices.
}
\vspace{-0.35cm}
\label{fig:reduction_overhead}
\end{figure}

\subsection{Component Impact Study}
\subsubsection{Reduction Operation}

We evaluated DAF's acceleration of reduction operators in Dynamic Quantization Control using conv2d and attention layers. For conv2d, the input tensor had a shape of [1, 256, 14, 14]. For attention operation, the input tensor was [1, 512, 768]. As illustrated in Figure~\ref{fig:reduction_overhead}, across multiple devices, DAF outperforms both the original training methods and NVIDIA CUB~\cite{CUB}, achieving speedups ranging from $11.2\times$ to $41.0\times$.



\vspace{-0.1in}
\begin{table}[!t]
    \center
    \small
    \caption{\small Comparison of different packing and unpacking methods. UL means Unified Layout.}
    \vspace{-0.1in}
    \begin{tabular}{|c|c|c|}
        \hline
         & Activation Memory & Training Time \\ 
         & + \textbf{Overhead} (MB) & + \textbf{Overhead} (ms) \\
        \hline
        Only CPU & 61+\textbf{32.1} & 323+\textbf{0} \\
        \hline
        Only GPU & 61+\textbf{0} & 323+\textbf{62} \\
        \hline
        CPU + GPU w/o UL & 61+\textbf{6.4} & 323+\textbf{51} \\
        \hline
        CPU + GPU w/ UL & 61+\textbf{6.4} & \textbf{282}+\textbf{60} \\
        \hline
    \end{tabular}
    \label{table:packing-unpacking}
\end{table}

\begin{table}[!t]
    \center
    \small
    \caption{\small Performance comparison of original methods and Red-Black Tree on different devices.}
    \vspace{-0.1in}
    \begin{tabular}{|c|c|c|}
        \hline
         & Original Method & Red-Black Tree  \\ 
        &  Time (us) &  Time (us) \\
        \hline
        Jetson AGX Orin & 1828 & 506 (\textbf{3.61\(\times\)})\\
        \hline
        Jetson AGX Xavier & 4680 & 1310 (\textbf{3.57\(\times\)}) \\
        \hline
        IQOO Neo3 & 17350  & 4901 (\textbf{3.54\(\times\)}) \\
        \hline
    \end{tabular}
    \label{table:red-black-tree}
\end{table}

\subsubsection{Packing and Unpacking}
To evaluate the efficiency of packing and unpacking operations, we conducted experiments on the Jetson AGX Xavier while training ResNet-18 on CIFAR-10. We compared four configurations: CPU-only, GPU-only, CPU and GPU collaboration without a unified layout, and CPU and GPU collaboration with a unified layout. The comparisons focused on both memory overhead and time overhead, where the time overhead refers to the duration of training a single iteration. 

For a fair comparison, the memory overhead is defined as the temporary memory usage incurred solely by the packing and unpacking operations. As shown in Table~\ref{table:packing-unpacking}, 
using the CPU and GPU collaboratively with a unified layout effectively reduces memory overhead while preserving nearly identical execution time performance. Moreover, collaborative packing with unified memory improves training time during the backward pass by unpacking activations in shared memory rather than DRAM, greatly reducing data transfer overhead between the GPU's global and shared memory.

\vspace{-0.2cm}
\subsubsection{Memory Management.}

We evaluated the performance of red-black tree for accelerating memory management and assessed DAF's ability to handle dynamic memory budgets. Table~\ref{table:red-black-tree} highlights the time differences per iteration between red-black tree-based management and a standard sorting-based approach across various devices. The results demonstrate that red-black tree management delivers approximately a $3.5\times$ speedup compared to the original method across all tested platforms.


\begin{figure}[!t]
\centering
{\includegraphics[width=1\linewidth]{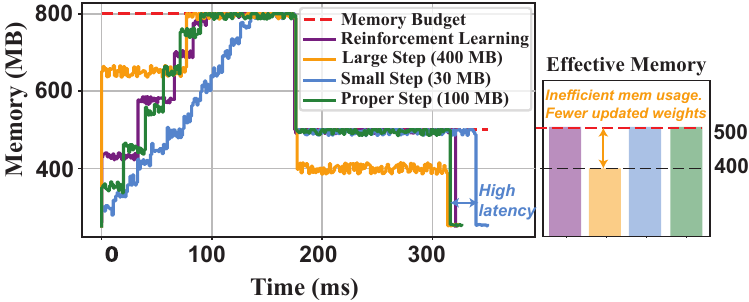}}
\vspace{-0.4cm}
\caption{\small Dynamic memory budget and DAF strategy on Jetson AGX Xavier with ResNet-18 for one training iteration.
}
\vspace{-0.2cm}
\label{fig:dynamic_mem_budget}
\end{figure}

\begin{figure}[!t]
\centering
{\includegraphics[width=1\linewidth]{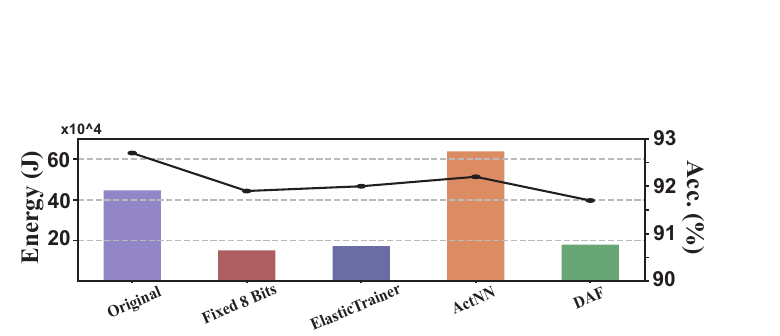}}
\vspace{-0.4cm}
\caption{\small Energy consumption comparison of ResNet-18 training on Jetson AGX Xavier.
}
\vspace{-0.2cm}
\label{fig:energy_resnet18}
\end{figure}

Moreover, Figure~\ref{fig:dynamic_mem_budget} compares the effectiveness of various strategies under a dynamic memory budget. We adopted the reinforcement learning-based method from DeepRM~\cite{mao2016resource} as our baseline and evaluated three memory increment strategies, corresponding to small (30 MB), proper (100 MB), and large (400 MB) allocation steps. The effective memory refers to the actual amount of memory utilized following a sudden reduction in the memory budget during a single training iteration. As illustrated, reinforcement learning, small-step, and proper-step approaches can adaptively manage memory to preserve activations efficiently. In contrast,  since memory can only be freed in units of entire allocated blocks, the large-step approach allocates memory in larger chunks, forcing the eviction of more activations than necessary. This excessive eviction leads to fewer weights being updated within that iteration, thereby prolonging overall training time. Meanwhile, the small-step strategy suffers from significant overhead due to frequent small and fragmented memory allocations. Among these, the proper-step strategy (100 MB) achieves performance comparable to reinforcement learning, with the additional advantages of eliminating extra training complexity and reducing decision-making overhead inherent to reinforcement learning.



\subsection{Energy Consumption}

On-device training typically consumes considerable energy. As illustrated in Figure~\ref{fig:energy_resnet18}, the original training method and ActNN exhibit significantly higher energy usage compared to other methods. In contrast, the energy usage of DAF is comparable to that of Fixed 8 Bits and ElasticTrainer. Although DAF introduces additional CPU computations, its kernel optimizations and shorter overall training duration ensure competitive total energy efficiency.

%% file: 5-Future_Work.tex

\section{Future Work}\label{sec:future_work}

Further research could explore heterogeneous collaboration to enhance performance beyond DAF's current capabilities. By carefully orchestrating task pipelines and offloading partial computations to idle processing units, it may achieve additional speedups.


%% file: 6-Conclusion.tex
\section{Conclusion}\label{sec:conclusion}
In this paper, we introduce DAF, a Dynamic Activation Framework designed to enable scalable and efficient on-device training. DAF offers system-level support for diverse dynamic activation quantization training algorithms, enabling practical memory savings and performance improvements. We have conducted extensive experimental evaluations across a variety of settings and observed that DAF can achieve up to $22.9\times$ memory savings and $3.2\times$ speedup on various edge and mobile platforms.

%% file: 7-DISCLAIMER.tex
\section{Disclaimer}
This paper was prepared for information purposes by the teams of researchers from the various institutions identified above, including the Global Technology Applied Research group of JPMorgan Chase Bank, N.A.. This paper is not a product of the Research Department of JPMorgan Chase Bank, N.A. or its affiliates. Neither JPMorgan Chase Bank, N.A. nor any of its affiliates make any explicit or implied representation or warranty and none of them accept any liability in connection with this paper, including, but limited to, the completeness, accuracy, reliability of information contained herein and the potential legal, compliance, tax or accounting effects thereof. This document is not intended as investment research or investment advice, or a recommendation, offer or solicitation for the purchase or sale of any security, financial instrument, financial product or service, or to be used in any way for evaluating the merits of participating in any transaction.